\documentclass[aps,prb,twocolumn,groupedaddress,showpacs,floatfix]{revtex4}
\usepackage{graphicx}
\begin{document}
\pagenumbering{arabic}
\pagestyle{plain}
\title{Distribution of Non-uniform Demagnetization Fields in Paramagnetic Bulk Solids}  
\author{Ross Dickinson,$^{1}$ A. Timothy Royappa,$^{2}$ Florentina Tone,$^{3}$ Laszlo Ujj,$^{1}$ and Guoqing Wu$^{1}$}
\affiliation{$^{1}$Department of Physics, University of West Florida, Pensacola, FL 32514, USA}
\affiliation{$^{2}$Department of Chemistry, University of West Florida, Pensacola, FL 32514, USA}
\affiliation{$^{3}$Department of Mathematics, University of West Florida, Pensacola, FL 32514, USA}
\altaffiliation{Corresponding Author. E-mail address: gwu@uwf.edu}
\date{\today}
\begin{abstract}
  A general calculation for the distribution of non-uniform demagnetization fields in paramagnetic bulk solids is described and the fields for various sample geometries are calculated. Cones, ellipsoids, paraboloids and hyperboloids with similar sample aspect ratios are considered. Significant differences in their demagnetization fields are observed. The calculation shows that the demagnetization field magnitudes decrease along the axis of symmetry (along $z$) where an externally applied magnetic field is aligned, and increase in the vicinity of the lateral surfaces with the largest field values found in the cone and the narrowest field distributions found in the hyperboloid. Application is made to the theoretical modeling of the $^{1}$H-NMR spectra of a single crystal of field-induced superconductor $\lambda$-(BETS)$_{2}$FeCl$_{4}$ with a rectangular sample geometry, providing a good fit to the measured NMR spectra. This calculation is also applicable to diamagnetic or ferromagnetic materials in general.     
\end{abstract}
\maketitle
\section{Introduction}
   The demagnetization field as a type of local magnetic field contributes to both an inhomogeneous broadening of nuclear magnetic resonance (NMR) spectra and a shift in the NMR resonance frequencies. Similar effects could be produced by other local field sources at a nucleus as well when they are not negligible.\cite{carter, slichter, abrag} Thus demagnetization fields are of particular interest in NMR spectroscopy and frequency shift (Knight shift) analysis, especially for samples with large magnetic susceptibility.\cite{carter, slichter, abrag, drain, dick} Unlike other local field sources, demagnetization fields are always associated with the sample geometries. Because the demagnetization field calculation is complex, a general calculation is rarely available in the literature. Instead, a common approach is to simplify the calculation by considering more manageable sample geometries like rectangles, cylinders and spheres and using the average of the field as represented by the so called ``demagnetization factor''.\cite{carter, osborn, joseph, aharoni, barr} The weakness here is that the demagnetization fields are usually not uniform throughout the sample even when the sample magnetization is spatially uniform.

  Another approach for the demagnetization field calculations is to use the fictitious ``magnetic charges'' as studied by Mozurkewich $et~ al.$,\cite{moz} Barbara,\cite{barbara} and Wanas $et~ al.$, \cite{wanas} in which samples with cylindrical geometries were again used. 
 
  In this paper, a general calculation for the distribution of non-uniform demagnetization fields in paramagnetic bulk solids is described. This method involves surface currents and volume currents originating from the sample magnetization when samples of various geometries are exposed to an externally applied magnetic field. Samples with cone, ellipsoid, paraboloid and hyperboloid geometries are investigated. Significant differences in their demagnetization fields are found, revealing the importance of the sample geometry for the field. Application is made to the theoretical modeling of the $^{1}$H-NMR spectra of a two-dimensional (2D) magnetic field-induced superconductor (FISC) $\lambda$-(BETS)$_{2}$FeCl$_{4}$ (single crystal), providing a good fit to the measured spectra.\cite{wu}              
\begin{figure}
\includegraphics[scale= 0.22]{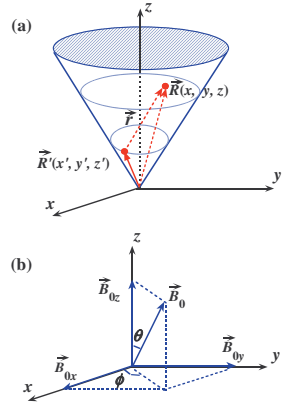}
\caption{(color online) (a) Cartesian coordinate system for a sample, where ($\it{x, y, z}$) is the field point inside (or outside) the sample volume, and ($\it{x', y', z'}$) is the source point (such as the point on the surface). (b) The orientation of the applied magnetic field $\vec{B_{0}}$. \label{fig1}}  
\end{figure}  
%
%
\section{Demagnetization field calculation}
  The demagnetization field ($\vec{B}$) originates from the magnetic polarization of particles. The polarization creates magnetic dipoles and forms a sample magnetization ($\vec{M}$) that leads to a macroscopic surface current density ($\vec{j}$) on the sample surface and a volume current density ($\vec{J}$) (if $\vec{M}$ is not uniform) inside the sample volume.   

  Any current produces its own magnetic field in space, not excepting for the surface current (density $\vec{j}$) and volume current (density $\vec{J}$) arising from the sample magnetization $\vec{M}$, certainly. In any case, it always depends on the sample geometry. The demagnetization field $\vec{B}$ can be expressed by
\begin{equation}
   \vec{B}(x, y, z) = \frac{\mu_{0}} {4\pi} \int\!\!\!\!\!\int_{A^{'}}\!\!\!\!\!\!\!\!\!\!\!\bigcirc ~~\frac{\vec{j}\times \vec{r}} {r^{3}} dA'
+ \frac{\mu_{0}} {4\pi} \int\!\!\!\!\int\!\!\!\!\int_{V'} \frac{\vec{J} \times \vec{r}} {r^{3}} dV', \\
\end{equation}
where $\mu_{0}$ is the permeability of free space, $\it{A'}$ is the sample surface area, $\it{V'}$ is the volume space enclosed by the area $\it{A'}$, and $\vec{r}$ is the separation vector, as shown in Fig. 1 (a), from the source point ($\it{x', y', z'}$) to the field point ($\it{x, y, z}$) in space, i.e., 
\begin{equation}
    \vec{r} =  (x - x')\hat{i} + (y - y')\hat{j} + (z - z')\hat{k}.
\end{equation}

  The first term in Eq. (1) is the field produced by the surface current density $\vec{j}$; it is a point-dependent closed-surface integral in a vector space. The second term in Eq. (1) is the field produced by the volume current density $\vec{J}$; it is a point-dependent volume integral in the vector space. Here $\vec{j}$ and $\vec{J}$ are 
\begin{eqnarray}
    \vec{j} =  \vec{M} \times \hat{n}, ~~~~~\text{and} ~~~~~\vec{J} = \nabla \times \vec{M}, 
\end{eqnarray}
respectively, where $\hat{n}$ is the unit vector normal to the sample surface, and $\nabla$ is the gradient operator.

  For paramagnetic materials (with non-interacting or weakly-interacting moments), the magnetization essentially has a very accurate linear dependence \cite {ash} on the externally applied magnetic field ($\vec{B}_0$), i.e.,
\begin{equation}
    \vec{M} = \chi \vec{H}_0,  ~~~~~\text{and}~~~~~\vec{H}_0 = \vec{B}_0/\mu_0,   \\
\end{equation}
where $\vec{H}_0$ is the applied magnetic field intensity, $\chi$ is the material's magnetic susceptibility (isotropic), and $\vec{B}_0$ = $B_{0x}\hat{i}$ + $B_{0y}\hat{j}$ + $B_{0z}\hat{k}$, as shown in Fig. 1 (b). Note, the error for the linear relation between $\vec{M}$ and $\vec{H}_0$ in Eq. (4) is negligible as it is in the order of $\sim$ $\chi^{2}\vec{H}_0$, where $\chi$ $<$ $\sim$ 10$^{-3}$ (cm$^{3}$/mol-ion) for most known materials. Moreover, if magnetic anisotropy is considered, then $M_{i}$ = $\chi_{ij} H_{0j}$, where $\chi_{ij}$ is the tensor element of $\chi$ ($\chi_{ij}$ = $\partial M_{i}$/$\partial H_{0j}$) and $i, ~j$ = $x, ~y, ~z$.

  As seen from Eq. (4), when a sample is placed in a homogeneous magnetic field $\vec{B}_0$, which is often the case as the spatial homogeneity of $\vec{B}_0$ is very necessary for NMR experiments, there is essentially no spatial variation in the magnetization throughout the sample volume $\it{V'}$. Thus the volume current $\vec{J}$ vanishes, i.e., the 2nd term in Eq. (1) $\vec{J} = \nabla \times \vec{M}$ = 0.  In other words, the current density only appears on the sample surface macroscopically. 

  For convenience of practical calculations using Eq. (1), we introduce a surface characteristic function $f(x', y', z')$ to describe the sample surface where $f(x', y', z')$ = 0 is satisfied [for example, for a unit sphere $f(x', y', z')$ = $x'^{2}$ + $y'^{2}$ + $z'^{2}$ $-$ 1 ]. Then the unit normal $\hat{n}$ for the surface and the surface element $dA'$ are 
\begin{equation}
  \hat{n} = \frac{\nabla f}{|\nabla f|},  ~~~~~\text{and}~~~~~dA' = \frac{|\nabla f|}{|(\nabla f)_{z'}|}dx'dy', 
\end{equation}
respectively, where $(\nabla f)_{z'}$ is the $z'$-component of the gradient of $f(x', y', z')$, i.e., $(\nabla f)_{z'}$ = $\partial f/\partial z'$. Note,  $dA'$ in Eq. (5) comes from $dA'$ = $dx'dy'/|\hat{n}\cdot\hat{k}|$. If $\hat{n}\cdot\hat{k}$ = 0, then $dA'$ = $dy'dz'/|\hat{n}\cdot\hat{i}|$ = $\frac{|\nabla f|}{|(\nabla f)_{x'}|}dy'dz'$ (if $\hat{n}\cdot\hat{i}$ $\neq$ 0), or $dA'$ = $dz'dx'/|\hat{n}\cdot\hat{j}|$ = $\frac{|\nabla f|}{|(\nabla f)_{y'}|}dz'dx'$ (if $\hat{n}\cdot\hat{j}$ $\neq$ 0).

  Thus by combining Eqs. (1)-(5), we have
\begin{equation}
   \vec{B}(x, y, z) = \frac{\chi} {4\pi} \int\!\!\!\!\!\int_{x'y'}\left(\vec{B}_0\times \frac{\nabla f}{|(\nabla f)_{z'}|} \right) \times \frac{\vec{r}}{r^{3}} dx'dy'. 
\end{equation}
  Equation (6) states that if the sample surface is describable by a surface characteristic function $f(x', y', z')$, then the demagnetization field $\vec{B}(x, y, z)$ can be calculated straightforwardly with a given external applied magnetic field $\vec{B}_0$. 
  
  In the following, sample geometries of cone, ellipsoid, paraboloid and hyperboloid are examined. 

  For example, for a right elliptical cone (0 $\leq$ $z'$ $\leq c$) the surface characteristic functions are $f(x', y', z')$ = $x'^2/a^2$ + $y'^2/b^2$ $-$ $z'^2/c^2$ and $f(x', y', z')$ = $c$ $-$ $z'$, for the lateral surface and the top radial surface (at $z'$ = $c$), respectively. Here $a$, $b$ and $c$ are the semimajor axes along the $x'(x)$, $y'(y)$ and $z'(z)$ axes, respectively. With Eq. (6), it gives the $x$, $y$, and $z$ components of the demagnetization field from the lateral surface current contributions as 
\begin{widetext}
\begin{eqnarray}
   B_{x} = \frac{\chi} {4\pi} \int\!\!\!\!\!\int_{x'y'}\frac{\left[\left(\frac{z'B_{0x}}{c^2} + \frac{x'B_{0z}}{a^2}\right)(z - z') - \left(\frac{y'B_{0x}}{b^2} - \frac{x'B_{0y}}{a^2}\right)(y - y')\right]} {\left[(x - x')^2 + (y - y')^2 + (z - z')^2 \right]^{3/2}} \cdot \frac{c^2}{|z'|} dx'dy', \\
   B_{y} = -\frac{\chi} {4\pi} \int\!\!\!\!\!\int_{x'y'}\frac{\left[\left(-\frac{z'B_{0y}}{c^2} - \frac{y'B_{0z}}{b^2}\right)(z - z') - \left(\frac{y'B_{0x}}{b^2} - \frac{x'B_{0y}}{a^2}\right)(x - x')\right]} {\left[(x - x')^2 + (y - y')^2 + (z - z')^2 \right]^{3/2}} \cdot \frac{c^2}{|z'|} dx'dy', \\
   B_{z} = \frac{\chi} {4\pi} \int\!\!\!\!\!\int_{x'y'}\frac{\left[\left(-\frac{z'B_{0y}}{c^2} - \frac{y'B_{0z}}{b^2}\right)(y - y') - \left(\frac{z'B_{0x}}{c^2} + \frac{x'B_{0z}}{a^2}\right)(x - x')\right]} {\left[(x - x')^2 + (y - y')^2 + (z - z')^2 \right]^{3/2}} \cdot \frac{c^2}{|z'|} dx'dy'. 
\end{eqnarray} 
\end{widetext}

  Similarly, the contributions from the current on the radial surface [$f(x', y', z')$ = $c$ $-$ $z'$] are 
\begin{widetext}
\begin{eqnarray}
  B_{x} = \frac{\chi} {4\pi} \int\!\!\!\!\!\int_{x'y'}\frac{B_{0x}(z - z')} {\left[(x - x')^2 + (y - y')^2 + (z - z')^2 \right]^{3/2}} dx'dy',  \\
  B_{y} = \frac{\chi} {4\pi} \int\!\!\!\!\!\int_{x'y'}\frac{B_{0y}(z - z')} {\left[(x - x')^2 + (y - y')^2 + (z - z')^2 \right]^{3/2}} dx'dy',  \\
  B_{z} = -\frac{\chi} {4\pi} \int\!\!\!\!\!\int_{x'y'}\frac{B_{0x}(x - x')+ B_{0y}(y - y')} {\left[(x - x')^2 + (y - y')^2 + (z - z')^2 \right]^{3/2}} dx'dy'.
\end{eqnarray} 
\end{widetext} 
Then the total demagnetization field $\vec{B}(x, y, z)$ is the vector summation of the two parts, from Eqs. (7)-(9) and from Eqs. (10)-(12). 

  Similar expressions to Eqs. (7) - (12) can be obtained for the ellipsoids, paraboloids and hyperboloids, or any other sample geometries as well by using their corresponding surface characteristic functions $f(x', y', z')$. 

  Figure 2 shows the calculated result for the demagnetization field magnitudes and directions for samples with hyperboloid and ellipsoid geometries with $\vec{B}_0$ along the $z$-axis (see Fig. 1). For simplicity the calculations were made by setting $a$ = $b$ = $c$ here. The data plots on the left side show the magnitude only (plotted as $H/M_{z}$ vs $x/a$, where $M_{z}$ is the sample magnetization along the $z$-axis). The vector plots on the right side include both the magnitudes and the directions. The corresponding results for samples with paraboloid and cone geometries are shown in Fig. 3.  
\begin{figure}
\includegraphics[scale= 0.44]{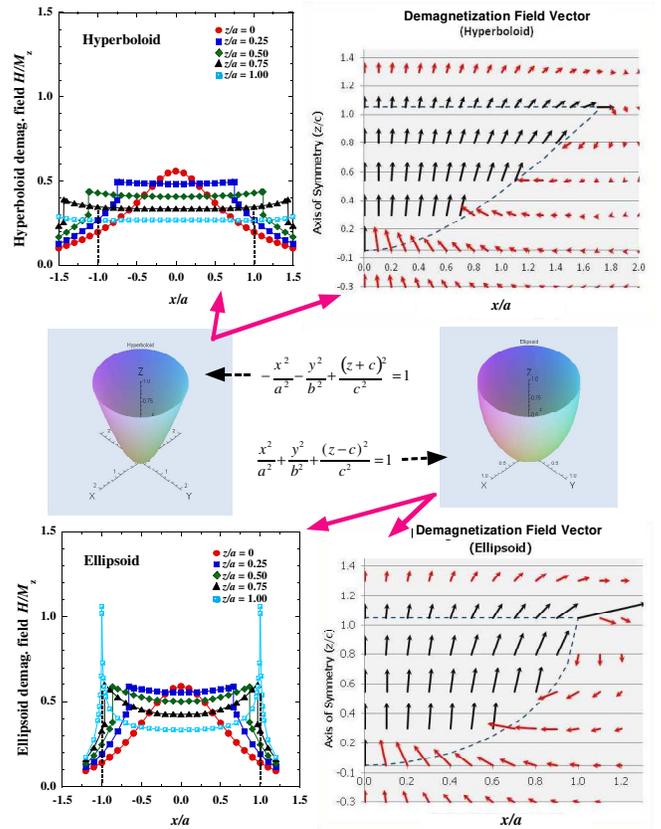}
\caption{(color online) Calculated demagnetization field magnitudes (data plots on the left side) and directions (vector plots on the right side) for samples with hyperboloid and ellipsoid geometries (shown in the middle), with applied magnetic field $\vec{B}_0$ along the $z$-axis. \label{fig2}}  
\end{figure}  
\begin{figure}
\includegraphics[scale= 0.44]{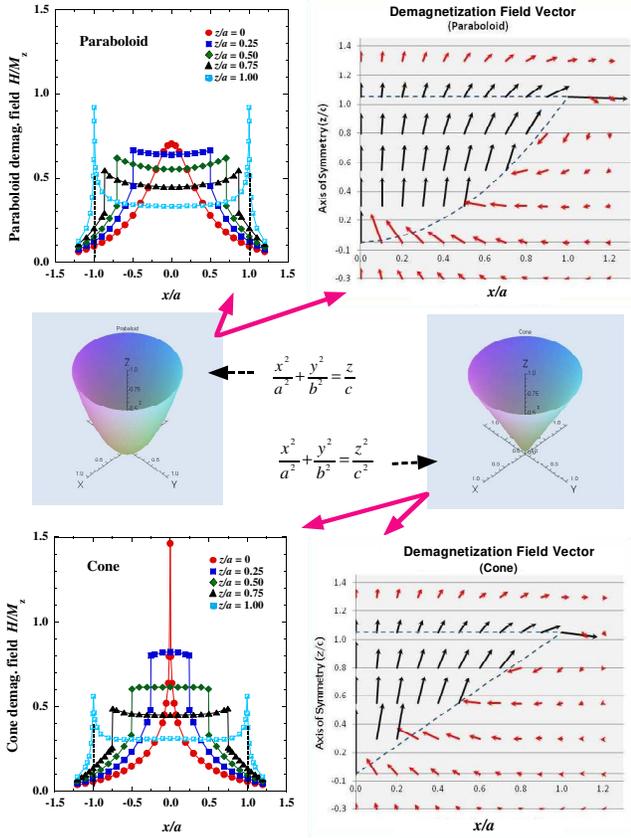}
\caption{(color online) Calculated demagnetization field magnitudes (data plots on the left side) and directions (vector plots on the right side) for samples with paraboloid and cone geometries (shown in the middle), with applied magnetic field $\vec{B}_0$ along the $z$-axis. \label{fig3}}  
\end{figure}  

  For comparison, the corresponding conditions in the calculation for $\vec{B}_0$, the $a$, $b$ and $c$ parameters for all the sample geometries, and the scales for the corresponding axes of all the plots in Figs. 2 - 3 are kept the same. We also define $c/a$ as the sample aspect ratio.    

  Figures 2 - 3 indicate that the demagnetization fields are non-uniform both inside and outside the samples. They also indicate that both the magnitude and the distribution of the demagnetization fields significantly depend on the sample geometries at a given $\vec{B}_{0}$ (along z). The data show that the demagnetization fields are generally smaller for the points closer to the $z$-axis, which is understandable even though the cone geometry is slightly different in the region of $z/a$ $\leq$ 0.5 (note, not all the data points shown here are close enough to the apex point or the lateral surface). The fields are also smaller along the direction of $\vec{B}_{0}$ since there is no current density ($\vec{j}$ = $\vec{M}\times\hat{n}$ = 0) at the radial surface (when $\vec{B}_0$ is aligned along z). There is a sharp discontinuity at the edges ($x/a$ = 1) except for the hyperboloid, and the integration diverges at the apex for the cone. Among them the smallest distribution (a general spatial variation) of the demagnetization fields inside the samples is in the hyperboloid under the similar parameter space. 

  For further investigation of the field distribution throughout the samples, we applied Eq. (6) for individuals with fixed geometries. 

  As an example, Fig. 4 shows the calculated result for the distribution of the demagnetization field $B_{z} (x, y, z)$ [the $z$-component of $\vec{B}(x, y, z)$ (along the direction of $\vec{B}_0$)] inside the samples with various sample aspect ratios $c/a$ under a cone geometry. The $z$-component $B_{z} (x, y, z)$ is of interest because it is the component that is along the external magnetic field direction that actually contributes to the NMR Hamiltonian (a negative dot product of the magnetic moment and external field).\cite{carter, slichter, wu} The line shapes are calculated from Eqs. (7)-(9) and use a Gaussian of width (FWHM) $2\sqrt{2ln2}~\delta$, where $\delta$ = $\sqrt{10}$ (arb. unit). The vertical scales are normalized for comparison. Similar results can be obtained using Eq. (6) for samples with ellipsoid, paraboloid, and hyperboloid geometries (not shown here).     
\begin{figure}
\includegraphics[scale= 0.29]{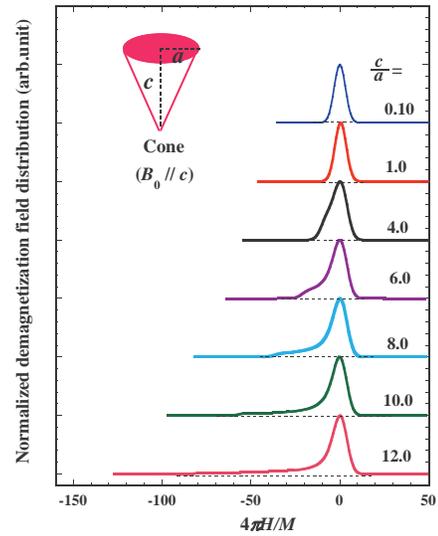}
\caption{(color online) Calculated demagnetization field distribution for samples with a cone geometry for various sample aspect ratios $c/a$, with applied magnetic field $\vec{B}_0$ along the $z$-axis ($B_{0}~ //~ c$). \label{fig4}}  
\end{figure}  

  Interestingly, Fig. 4 shows that in viewing along the applied magnetic field $\vec{B}_0$ direction, the demagnetization field distribution (corresponding to the NMR absorption spectrum) is the narrowest and has essentially no difference for samples either very thin (small $c/a$) or very thick (large $c/a$; long sample). This observation is also true for samples with a cylindrical geometry,\cite{moz} suggesting its suitability for samples with other geometries as well. But it does have a strong dependence on the sample aspect ratio $c/a$. 

  When the sample aspect ratio is small ($c/a$ $\leq$ 1), the demagnetization field variation in space along the direction of the applied magnetic field $\vec{B}_{0}$ is narrow, while the major field variation actually happens most strongly in the direction that is perpendicular to $\vec{B}_0$. This can be seen from Figs. (2)-(3) (case $c/a$ = 1), where the demagnetization field variation along $x$ (the radial direction) is apparently larger (see the red curves) than that along $z$ (view along $z$ from $z/a$ = 0 to $z/a$ = 1). This is also true for all other geometries. 

  As the sample aspect ratio $c/a$ increases, Fig. 4 shows that the demagnetization field distribution inside the sample gets broader with the appearance of a left shoulder (i.e, the line is broadened). This happens starting at $c/a$ $\sim$ 1. The shoulder is significant in the range $c/a$ $\sim$ 3 $-$ 12. As $c/a$ further increases, the shoulder is diminished gradually while it gets broader and broader, and it becomes insignificant at $c/a$ $\geq$ $\sim$ 20. However, there is little change in the major peak, and the line essentially becomes a single narrow peak again.
\section{discussion}      
  In this section, we have more comparison discussions regarding the sample geometries used in the calculations, and we also discuss the result of the application for the theoretical modeling of an NMR spectrum.    

  Figure 5 shows a typical comparison for the calculated demagnetization field magnitude for samples with cone, ellipsoid, paraboloid and hyperboloid geometries in their middle planes at $z/a$ = 0.5 ($c/a$ = 1), with applied magnetic field $\vec{B}_0$ along the $z$-axis. The data comes from a set of data shown in Figs. 2 - 3 (the green curves). Figure 5 indicates that among them the demagnetization field in the cone is the largest, and the smallest demagnetization field is in the hyperboloid. Samples with a hyperboloid geometry also have the smallest distribution of demagnetization field under similar sample aspect ratio $c/a$.  

  In comparison with other geometries, such as the cylindrical ones,\cite{moz, barbara} with similar sample aspect ratio ($c/a$ = 1, for example) and the same $\vec{B}_0$ [including the field alignment] the demagnetization field magnitude inside a cylindrical sample is also smaller than that for samples with a cone geometry correspondingly (like that in a $z/a$ = 0.5 plane). But the demagnetization field distribution in cylindrical geometries is apparently larger and it is also more sensitive to the sample aspect ratio $c/a$.\cite{moz} A more detailed comparison can be made directly by the calculation using the similar forms as Eq. (6), i.e., 
\begin{equation}
\vec{B}(x, y, z) = \frac{\chi} {4\pi} \int\!\!\!\!\!\int_{z'x'}\left(\vec{B}_0\times \frac{\nabla f}{|(\nabla f)_{y'}|} \right) \times \frac{\vec{r}}{r^{3}} dz'dx',
\end{equation}
\begin{equation}
\vec{B}(x, y, z) = \frac{\chi} {4\pi} \int\!\!\!\!\!\int_{y'z'}\left(\vec{B}_0\times \frac{\nabla f}{|(\nabla f)_{x'}|} \right) \times \frac{\vec{r}}{r^{3}} dy'dz'. 
\end{equation}

  For samples with a cubic geometry, we made similar calculations using Eqs. (13) - (14) and found that their demagnetization field magnitudes at the $z/a$ = 0.5 plane were $\sim$ 10$\%$ larger than those of the cone (here $a$ is the cube side length) with the same $\vec{B}_0$, while the field distribution inside the cube was the narrowest among all.
\begin{figure}
\includegraphics[scale= 0.23]{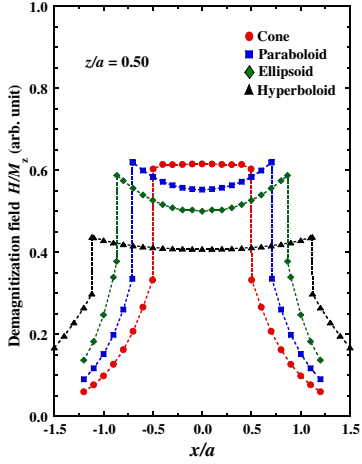}
\caption{(color online) Calculated demagnetization field magnitudes for samples with cone, ellipsoid, paraboloid and hyperboloid geometries in the middle plane $z/a$ = 0.5, with applied magnetic field $\vec{B}_0$ along the $z$-axis. Here the sample aspect ratio $c/a~=~1$. \label{fig5}}  
\end{figure}
\begin{figure}
\includegraphics[scale= 0.23]{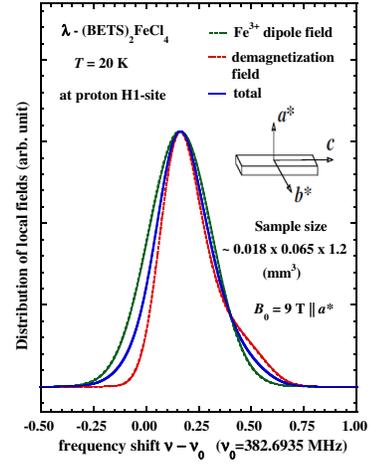}
\caption{(color online) Calculated distributions of local magnetic fields at a typical proton (H1) site \cite{wu} in a single crystal of $\lambda$-(BETS)$_{2}$FeCl$_{4}$ at 20 K and $B_{0}$ = 9 T. The dashed (red) line is for the demagnezation field, the solid (green) line is for the Fe$^{3+}$ dipole field, and the solid (blue) curve shows their normalized total. \label{fig6}}  
\end{figure}  
\begin{figure}
\includegraphics[scale= 0.30]{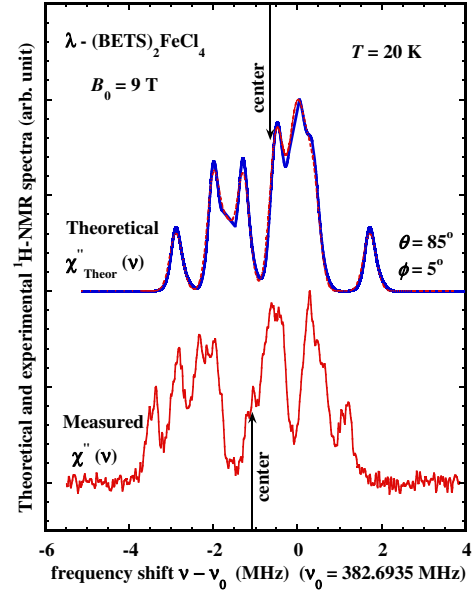}
\caption{(color online) Comparison of the calculated $^{1}$H-NMR absorption spectrum with the measured spectrum of a single crystal of $\lambda$-(BETS)$_{2}$FeCl$_{4}$ at 20 K, for magnetic field $B_{0}$ = 9 T aligned with $\theta$ = +85$^{\circ}$ and $\phi$ = +5$^{\circ}$. The dashed (red) line is the result \cite {wu} that did not include the distribution of demagnetization field contribution at the proton sites, while the solid (blue) line is the result that does it. \label{fig7}}  
\end{figure}  

  Figure 6 shows the calculated distribution of the demagnezation field using Eqs. (13)-(14) as compared with the distribution of Fe$^{3+}$ dipole field at a typical proton site (H1 site),\cite {wu}in a single crystal of $\lambda$-(BETS)$_{2}$FeCl$_{4}$, considering the materials magnetic susceptibility and the needle-like sample used for the NMR experiment (at 20 K and at magnetic field $B_{0}$ = 9 T). The sample size is (1.2 $\pm$ 0.1)$\times$(0.065 $\pm$ 0.010)$\times$(0.018 $\pm$ 0.005) mm$^{3}$.
  
 The material $\lambda$-(BETS)$_{2}$FeCl$_{4}$ is paramagnetic (at 20 K and 9 T) and has a large magnetic susceptibility $\chi$ ($\chi$ $\sim$ 0.15 emu/mol.Oe) due to the high spin state of the 3d Fe$^{3+}$ ion electron moments (spin $S$ = 5/2). \cite{kob} The local magnetic fields and the field distributions are dominant by the dipolar field from the Fe$^{3+}$ ions, but they also have the contributions from the demagnetization field, Lorentz field, exchange field, proton-proton dipolar field, hyperfine couplings fields, etc.. Not all of these local field contributions are negligible. There are 16 different proton sites in each unit cell in this material. In fact, at some proton sites the magnitudes of the demagnetization fields are even larger than the dipolar fields of the Fe$^{3+}$ ions. This is also true for the Lorentz fields.\cite{wu} Thus, the demagnetization field in this material could have a significant contribution to the local magnetic field magnitude and maybe distributions as well, depending on the sample geometry and the $\vec{B}_0$ alignment.   

  Figure 7 shows the new result for the theoretical modeling of a NMR spectrum as compared with the measured spectrum of the single crystal of $\lambda$-(BETS)$_{2}$FeCl$_{4}$ at 20 K, for magnetic field $B_{0}$= 9 T aligned with $\theta$ = +85$^{\circ}$ and $\phi$ = +5$^{\circ}$, where $\theta$ and $\phi$ describe the direction of $\vec{B}_{0}$ relative to the crystal lattice. Earlier report \cite{wu} did not include the contribution of the distribution of the demagnization field for simplicity, while the new result does which has an application using Eq. (6) [surface currents appearing on all the surfaces of the sample] similar to that shown in Fig. 6 for all the protons sites in the crystal lattice. 

  Noticeably, even though the result shown in Fig. 7 indicates that the effect of the distribution of demagnetization field to the shape of the NMR spectrum is not significant in this case, but there is still significant $^{1}$H-NMR resonance frequency shift contributed by the demagnetization field. The shift is $\sim$ 45$\times$10$^{-4}$T$\times$42.5759 MHz/T $\approx$ 191.6 kHz (i.e., $\sim$ 0.5$\%$) at $\sim$ 50 K [note, it is a significant shift even if an NMR frequency shift (Larmor) is 0.1$\%$].\cite{slichter, abrag} The small differences between the calculated and measured spectra could mainly come from the $\pi$-$d$ interactions and $d$-$d$ interactions of the electron moments in the material,\cite{kob} which are not included in Eq. (6).

  On an atomic scale, there could be large spatial local field variations, thus Eq. (6) can be regarded as the average field over a macroscopic small volume $dV'$ which is large enough to contain many atoms (molecules) or a unit cell. Thus, in the spectrum modeling, the difference of demagnetization field at each proton site in a unit cell of the crystal lattice is not considered.   

  We have used paramagnetic samples for the demagnetization field calculations, but Eq. (6) can also be used for diamagnetic and ferromagnetic materials, as long as the linear isotropic relation $\vec{M}$ = $\chi \vec{H}_0$ in Eq. (4) applies (non-interacting  or weakly interacting moments). For diamagnetic materials, $\chi$ $<$ 0, while for ferromagnetic and paramagnetic materials $\chi$ $>$ 0. 
\section{Conclusions}
  In this study, a general calculation of the non-uniform demagnetization fields in paramagnetic bulk solids is described and the fields for various sample geometries are calculated. The calculations show significant differences in their non-uniform demagnetization fields, revealing the importance of the sample geometry to the field. Viewing from the externally applied magnetic field $\vec{B}_0$ direction when samples are either very thin (sample aspect ratio $c/a$ $\leq$ $\sim$ 0.1) or very long ($c/a$ $\geq$ $\sim$ 20), we observe that the demagnetization field distribution is narrow and in this case its contribution to the NMR absorption spectra line shape could be neglected. However, the demagnetization field magnitude and its contribution to the NMR frequency shift remain significant, depending on the sample magnetic susceptibility as evidenced by our application for the theoretical modeling of the NMR spectra of a single crystal of 2D magnetic field-induced superconductor $\lambda$-(BETS)$_{2}$FeCl$_{4}$. We suggest that this observation be used for all geometries.  
\begin{acknowledgments}
   We thank W. G. Clark and S. E. Brown at UCLA for the NMR measurements, which were supported by the NSF Grant No. DMR-0334869(W.G.C.) and 0520552 (S.E.B.). We also thank Leonard W. ter Haar and C. S. Prayaga at UWF for helpful discussions and support.  
\end{acknowledgments}

\end{document}